\documentclass{mem}
\usepackage{natbib}\usepackage{txfonts}\usepackage{balance}
\usepackage{graphicx}
\usepackage[a4paper,breaklinks,dvipdfm]{hyperref}
\idline{75}{282}

\begin{document}

\title{
Gamma-Ray Astronomy with ARGO-YBJ
}

   \subtitle{}

\author{
G. \,Di Sciascio on behalf of the ARGO-YBJ Collaboration
          }

  \offprints{G. Di Sciascio}

\institute{INFN -- Sez. di Roma Tor Vergata, Viale della Ricerca
Scientifica 1, I-00133 Roma, Italy\\ \email{disciascio@roma2.infn.it} }

\authorrunning{Di Sciascio }

\titlerunning{Gamma-Ray Astronomy with ARGO-YBJ}

\abstract{ARGO-YBJ is a \emph{full coverage} air shower array located at the YangBaJing Cosmic Ray Laboratory (Tibet, P.R. China, 4300 m a.s.l., 606 g/cm$^2$) recording data with a duty cycle $\geq$85\% and an energy threshold of a few hundred GeV. In this paper the latest results in Gamma-Ray Astronomy are summarized. 
  } \maketitle{}

\section{The ARGO-YBJ experiment}
%
The ARGO-YBJ detector is constituted by a central carpet large about 74$\times$78 m$^2$, made of a single layer of Resistive Plate Chambers (RPCs) with $\sim$93$\%$ of active area, enclosed by a guard ring
partially instrumented ($\sim$20$\%$) up to $\sim$100$\times$110
m$^2$. The apparatus has a modular structure, the basic data
acquisition element being a cluster (5.7$\times$7.6 m$^2$),
made of 12 RPCs (2.85$\times$1.23 m$^2$ each). Each chamber is
read by 80 external strips of 6.75$\times$61.80 cm$^2$ (the spatial pixels),
logically organized in 10 independent pads of 55.6$\times$61.8
cm$^2$ which represent the time pixels of the detector \cite{Aie06}. 
The read-out of 18360 pads and 146880 strips is the experimental output of the detector. 
The central carpet contains 130 clusters (hereafter ARGO-130) and the
full detector is composed of 153 clusters for a total active
surface of $\sim$6700 m$^2$.
All events giving a number of fired pads N$_{pad}\ge$ N$_{trig}$ in the
central carpet within a time window of 420 ns are recorded.
The spatial coordinates and the time of any fired pad are used to reconstruct the position of the shower core and the arrival direction of the primary, as
described in \citep{DiS07}.
The whole system, in smooth data taking since July 2006 firstly with ARGO-130, is in stable data taking with the full apparatus of 153 clusters since November 2007 with the trigger condition N$_{trig}$ = 20 and a duty cycle $\geq$85\%. The trigger rate is $\sim$3.5 kHz with a dead time of 4$\%$.

The performance of the detector (angular resolution, pointing accuracy, energy calibration) and the operation stability are monitored on a monthly basis by observing the Moon shadow, i.e., the deficit of CR detected in its direction. 
The measured angular resolution is better than 0.5$^{\circ}$ for CR-induced showers with energies E $>$5 TeV improving up to about 0.3$^{\circ}$ for E $>$10 TeV. The overall absolute pointing accuracy is $\sim$0.1$^{\circ}$.
The absolute pointing of the detector is stable at a level of 0.1$^{\circ}$ and the angular resolution is stable at a level of 10\%.
The absolute rigidity scale uncertainty of ARGO-YBJ is estimated to be less than 13\% in the range 1 - 30 TeV/Z \cite{DiS11}.

\section{Data analysis}

The dataset for the analysis presented in this paper contains all showers with zenith angle $\theta<$50$^{\circ}$ collected from November 2007 to February 2011. The total effective observation time is 1024 days and the total number of events is 1.7$\times$10$^{11}$.

A sky map in celestial coordinates (right ascension and declination) with 0.1$^{\circ}\times$0.1$^{\circ}$ bin size is filled with the detected events. 
The number of cosmic ray background events in each bin is calculated with the so-called ``direct integral method'' \cite{fleysher04}. The Li\&Ma formula is used to estimate the statistical significance of the observation in standard deviations (s.d.).

\section{Results}
%
Four known VHE $\gamma$-ray sources have been detected by ARGO-YBJ with a statistical significance greater than 5 s.d., i.e. Crab Nebula, Mrk 421, MGRO J1908+06 and MGRO J2031+41.

\subsection{Crab Nebula}

\begin{figure}[]
\resizebox{\hsize}{!}{\includegraphics[clip=true]{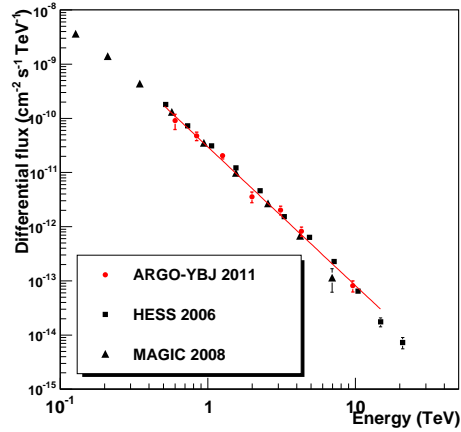}}
 \caption{\footnotesize Crab Nebula energy spectrum measured by ARGO-YBJ, compared with other measurements.}
\label{crab}
\end{figure}

With all data ARGO-YBJ observed a TeV signal from the Crab Nebula with a statistical significance greater than 17 s.d., proving that the cumulative sensitivity of the detector reached 0.3 Crab units level.
Assuming a power law spectrum, the obtained best fit in the energy range $\sim$0.5-10 TeV is (TeV$^{-1}$ cm$^{-2}$ s$^{-1}$):
 \begin{equation}
\frac{dN}{dE}=(3.0\pm0.30)\cdot 10^{-11} \left(\frac{E}{1\>TeV}\right)^{-2.59\pm0.09}
\end{equation}
in good agreement with other observations (Fig. \ref{crab}).
The quoted errors are purely statistical. We evaluate a systematic error on the flux less than 30$\%$ mainly due to the background 
evaluation and to the uncertainty on the absolute energy scale.
%
%

According to the AGILE and Fermi data \cite{agile2,fermi2,agile3} three major flaring episodes at energies E $>$100 MeV occurred during the ARGO-YBJ data acquisition. 

{\bf Flare 1}: starting time MJD 54857, duration $\Delta$t$\sim$16 days, 
maximum flux F$_{max}$ $\sim$5 times larger than the steady flux
\cite{fermi2}. During this flare no excess is present in our data, for any multiplicity threshold.

{\bf Flare 2}: starting time MJD 55457, duration $\Delta$t $\sim$4 days, 
maximum flux F$_{max}$ $\sim$5 times larger than the steady flux \cite{agile2, fermi2}. According to data analysis the $\gamma$-ray emission is concentrated in 3 narrow peaks of $\sim$12 hours duration each \cite{balbo,agile3}. Integrating the 3 transits we observe an excess of 3.1 s.d. for N$_{pad}>$40 over the expected steady flux (0.55 s.d.). 
If the excess were due to a flare, the $\gamma$-ray flux would be higher 
by a factor $\sim$5 with respect to the steady flux at energies around 1 TeV.
Integrating the data over 10 transits (from MJD 55456/57 to MJD 55465/66) the signal significance is 4.1 s.d. (pre-trial), while 1.0 s.d.
are expected from the steady flux \cite{argofla}. 
No measurement from Cherenkov telescopes are available in coincidence with our observations and with the different spikes to confirm this excess. Sporadic measurements performed by MAGIC and VERITAS telescopes at different times from MJD 55456.45 to MJD 55459.49 show no evidence for a flux variability \cite{magfla,verfla}.

{\bf Flare 3}: starting time MJD 55660 \cite{fermi3}, duration $\Delta$t $\sim$9 days, maximum flux F$_{max}$ $\sim$14 times larger than the steady flux
\cite{agile3}.
Integrating the ARGO-YBJ data over the 6 days in which AGILE detected a
flux enhancement, i.e. from MJD 55662.00 to MJD 55668.00, we report evidence for an excess for events with N$_{pad}>$ 100 at a level of about 3.5 s.d. . No measurements from Cherenkov telescopes are available during these days,
due to the presence of the Moon during the Crab transits.

\subsection{Mrk421}

Mrk421 was the first source detected by ARGO-YBJ in July 2006 when the detector started recording data with ARGO-130 only and was in commissioning phase.
ARGO-YBJ has monitored Mrk421 for more than 3 years above 0.3 TeV, studying the correlation of the TeV flux with X-ray data. We observed this source with a total significance of about 14 s.d., averaging over quiet and active periods. As it is well known, this AGN is characterized by a strong flaring activity both in X-rays and in TeV $\gamma$-rays. Many flares are observed in both X-ray and $\gamma$-ray bands simultaneously. The $\gamma$-ray flux observed by ARGO-YBJ has a clear correlation with the X-ray flux. No lag
between the X-ray and $\gamma$-ray photons longer than 1 day is found. The evolution of the spectral energy distribution is investigated by measuring spectral indices at four different flux levels. Hardening of the spectra is observed in both X-ray and $\gamma$-ray bands. The $\gamma$-ray flux increases quadratically with the simultaneously measured X-ray flux.
All these observational results strongly favor the synchrotron self-Compton process as the underlying radiative mechanism.
The results of Mrk421 long-term monitoring are summarized in \cite{bartoli11}

\subsection{MGRO J1908+06}

The $\gamma$-ray source MGRO J1908+06 was discovered by the MILAGRO \cite{mila07} at a median energy of $\sim$20 TeV and confirmed by HESS \cite{hess09} at energies above 300 GeV.
The Milagro and HESS energy spectra are in disagreement at a level of 2-3 s.d., being the Milagro result about a factor 3 higher at 10 TeV \cite{mila09b}.
%
\begin{figure}[]
\resizebox{\hsize}{!}{\includegraphics[clip=true]{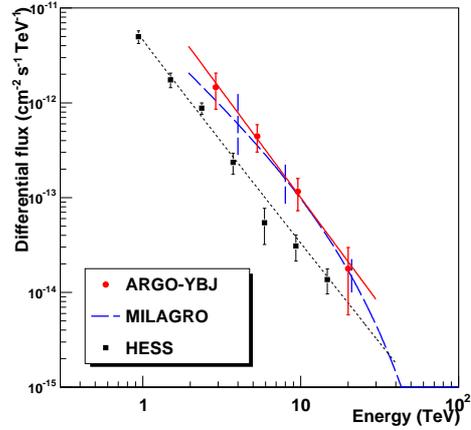}}
 \caption{\footnotesize MGRO J1908+06 energy spectrum measured by ARGO-YBJ compared to other experiments. The continuous line is the best fit to data. The plotted errors are purely statistical for all the detectors.}
\label{mj1908}
\end{figure}
%

ARGO-YBJ observed a TeV emission from MGRO J1908+06 with a significance of about 6 s.d.. The intrinsic extension is determined to be $\sigma_{ext}$ = 
0.50$^{\circ}$$\pm0.35$, consistent with the HESS measurement ($\sigma_{ext}$ = 0.34$^{\circ}$$_{-0.03}^{+0.04}$).
The best fit power law spectrum is (TeV$^{-1}$ cm$^{-2}$ s$^{-1}$) (Fig. \ref{mj1908}): 
 \begin{equation}
\frac{dN}{dE}=(2.2\pm0.4)\cdot 10^{-13} \left(\frac{E}{7\>TeV}\right)^{-2.3\pm0.3}
\end{equation}
Beside the statistical errors, this measurement could be affected by
a systematic uncertainty less than 30$\%$ mainly due to the background evaluation and to the determination of the absolute energy scale.

The spectrum is in agreement with Milagro, but only marginally consistent with HESS, being the ARGO-YBJ flux a factor $\sim$3 larger. 
The origin of such a disagreement could be a statistical fluctuation and/or the effect of systematic uncertainties (HESS reports a systematic error of $\sim$20$\%$ on the flux). 
However, since a higher flux has been observed also by Milagro, 
a possibility is that the larger flux is the consequence of the contamination of other extended sources lying near the object observed by HESS, 
since the former detectors have a worse angular resolution and integrate
the signal over a larger area.

\subsection{MGRO J2031+41}

MGRO J2031+41, located inside the Cygnus region, is the third significant source discovered by Milagro.
%
\begin{figure}[]
\resizebox{\hsize}{!}{\includegraphics[clip=true]{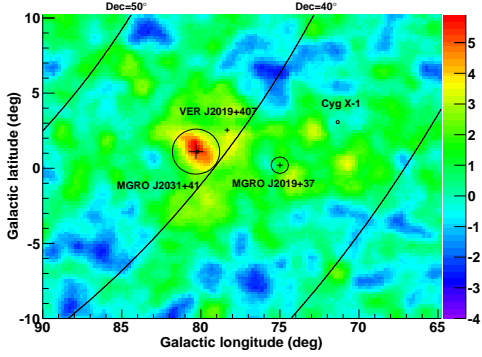}}
 \caption{\footnotesize ARGO-YBJ significance map of the Cygnus region. The four known VHE $\gamma$-ray source are shown.}
\label{cygno}
\end{figure}
%
\begin{figure}[]
\resizebox{\hsize}{!}{\includegraphics[clip=true]{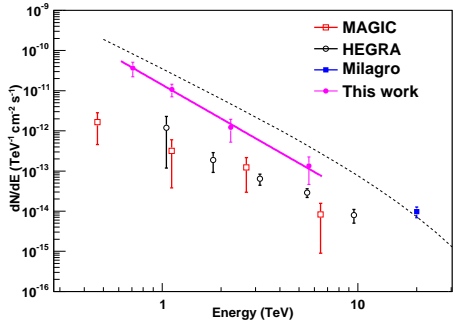}}
 \caption{\footnotesize MGRO J2031+41 energy spectrum measured by ARGO-YBJ compared with other experiments. The continuous line is the best fit to data. The dashed line indicates the Crab Nebula spectrum.}
\label{mj2031}
\end{figure}
%
Figure \ref{cygno} shows the ARGO-YBJ significance map of the Cygnus region. An excess is observed over a large part of the Cygnus region, which indicates a possible diffuse $\gamma$-ray emission. 

The highest significance value is 5.8 s.d. at (307.8$^{\circ}$, 41.9$^{\circ}$), consistent with the position of VHE sources MGRO J2031+41 and TeV J2032+4130.
The intrinsic extension of MGRO J2031+41 is determined to be $\sigma_{ext}$ = (0.2$_{-0.2}^{+0.4})^{\circ}$, consistent with the estimation by the MAGIC and HEGRA experiments. Assuming an intrinsic extension $\sigma_{ext}$ = 0.1$^{\circ}$, we estimated that the flux (TeV$^{-1}$ cm$^{-2}$ s$^{-1}$) in the energy range from 0.6 TeV to 7 TeV is
 \begin{equation}
\frac{dN}{dE}=(1.40\pm0.34)\cdot 10^{-11} \left(\frac{E}{1\>TeV}\right)^{-2.8\pm0.4}
\label{eq:mgro2031}
\end{equation}
which is 10 and 17 times higher than the flux of TeV J2032+4130 as determined by HEGRA and MAGIC, respectively, at energies above 1 TeV. The flux reported by Milagro at 20 TeV is also much higher than the naive extrapolation of the spectrum of TeV J2032+4130, as clearly shown in Fig. \ref{mj2031}.
Besides the statistical error shown in Eq. \ref{eq:mgro2031}, the systematic error in the flux determination is estimated to be less than 30\%. To estimate the contribution from the diffuse emission, a region of 3$^{\circ}$-5$^{\circ}$ around the source is adopted. The contribution to flux is about 25\% and depends on energy, being higher at lower energies. As a consequence, the spectral index seems softer.

\subsection{MGRO J2019+37}

MGRO J2019+37 was discovered by Milagro at 20 TeV \cite{abdo07a} with a statistical significance of about 12 s.d., thus resulting the most significant source in their data set after the Crab Nebula. The extension is estimated to be 0.32$^{\circ}$$\pm$0.12$^{\circ}$. The spectrum of this source seems to be hard with index -1.83 and cutoff at 22.4 TeV \cite{mila09b}. 

No excess above 3 s.d. is detected by ARGO-YBJ in the MGRO J2019+37 position. 
%
\begin{figure}[]
\resizebox{\hsize}{!}{\includegraphics[clip=true]{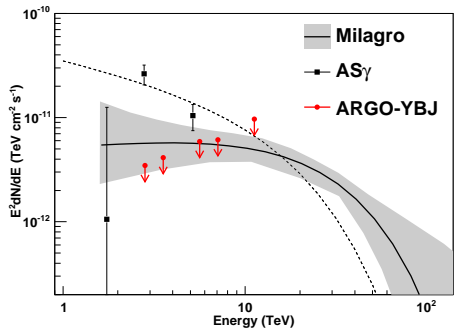}}
 \caption{\footnotesize ARGO-YBJ upper limits to the MGRO J2019+37 flux. The solid line shows the energy spectrum measured by Milagro. The dashed line shows the Crab Nebula spectrum.}
\label{mj2019}
\end{figure}
%
We set flux upper limits at 90\% confidence level (c.l.) (Fig. \ref{mj2019}) assuming the Milagro results for energy spectrum and extension. We note that the upper limits around 3 TeV exclude the Milagro flux.
For comparison, the high flux measured by Tibet AS$\gamma$ \cite{wang07} is also shown (square points).

\section{Conclusions}

The ARGO-YBJ experiment is an air shower array with large field of view and has been continuously monitoring the northern sky since November 2007. With all data recorded up to February 2011, 4 sources have been observed with a statistical significance greater than 5 s.d.: Crab Nebula, Mrk421, MGRO J1908+06 and MGRO J2031+41.

ARGO-YBJ has been continuously monitoring the Crab Nebula for more than 3 years. 
The measured energy spectrum in the energy range 500 GeV - 10 TeV is in agreement with the results of other experiments. No flux variations with a statistical significance larger than 5 s.d. have been detected in time scales of days or months. Nevertheless, enhanced flux of significance about 3 s.d. has been observed in coincidence with the occurrence of two flares detected by AGILE and Fermi. 

In the last 3 years we carried out a detailed long-term monitoring of Mrk421 detecting a number of TeV $\gamma$-ray flares correlated with X-ray flaring activity.

The energy spectra of MGRO J1908+06 and MGRO J2031+41 have been measured.
No signal from MGRO J2019+37 is detected, and the derived upper limits at 90\% c.l. are lower than the Milagro flux at energies below 3 TeV, indicating that the source flux may be not stable.

\bibliographystyle{aa}

\end{document}